\documentclass[aps,prb,twocolumn,showpacs,amsmath,amssymb]{revtex4-1}
\usepackage{graphicx}
\usepackage{dcolumn}
\usepackage{epstopdf}
\usepackage{bm}

\usepackage{color}
\usepackage[normalem]{ulem}

\begin{document}
\title{First-principles modelling of magnetic excitations in Mn$_{12}$}
\author{V.V. Mazurenko$^{1}$, Y.O. Kvashnin$^{2,3}$,  Fengping Jin$^{4}$, H.A. De Raedt$^{5}$, A.I. Lichtenstein$^{6}$, M.I. Katsnelson$^{7,1}$}
\affiliation{$^{1}$Theoretical Physics and Applied Mathematics Department, Ural Federal University, Mira Str.19,  620002
Ekaterinburg, Russia \\
$^{2}$ European Synchrotron Radiation Facility, 6 Rue Jules Horowitz, BP220, 38043 Grenoble Cedex, France \\
$^{3}$ Department of Physics and Astronomy, Division of Materials Theory, Uppsala University, Box 516, SE-75120 Uppsala, Sweden \\
$^{4}$ Institute for Advanced Simulation, J\"ulich Supercomputing Centre, Forschungszentrum J\"ulich, D-52425 J\"ulich, Germany \\
$^{5}$ Zernike Institute for Advanced Materials, University of Groningen, Groningen, The Netherlands\\
$^{6}$ Institute of Theoretical Physics, University of Hamburg, Jungiusstrasse 9, 20355 Hamburg, Germany\\
$^{7}$ Radboud University of Nijmegen, Institute for Molecules and Materials, Heijendaalseweg 135, 6525 AJ Nijmegen, The Netherlands}
\date{\today}

\begin{abstract}
We have developed a fully microscopic theory of magnetic properties of the prototype molecular magnet Mn$_{12}$. First, the intra-molecular magnetic properties have been studied by means of first-principles density functional-based methods, with local correlation effects being taken into account within the local density approximation plus $U$ (LDA+$U$) approach. Using the magnetic force theorem, we have calculated the interatomic isotropic and anisotropic exchange interactions and full tensors of single-ion anisotropy for each Mn ion. Dzyaloshinskii-Moriya (DM) interaction parameters turned out to be unusually large, reflecting a low symmetry of magnetic pairs in molecules, in comparison with bulk crystals.  Based on these results we predict a distortion of ferrimagnetic ordering due to DM interactions. Further, we use an exact diagonalization approach  allowing to work with as large Hilbert space dimension as 10$^8$ without any particular symmetry (the case of the constructed magnetic model). Based on the computational results for the excitation spectrum, we propose a distinct interpretation of the experimental inelastic neutron scattering spectra.
\end{abstract}
\maketitle

\section {Introduction}
Molecular magnets, such as Mn$_{12}$, Fe$_{8}$, Mn$_{4}$, V$_{15}$ are in the focus of modern science due to their potential for novel technologies such as molecular electronics,  solar-energy harversting, thermoelectrics, sensing and others. \cite{Bogani, Mannini} The functionality of molecular nanomagnets as materials for advanced technologies is mainly related to the control and manipulation of excited quantum spin states, which ultimately requires the microscopic identification of the total spin, energies and lifetimes corresponding to different magnetic excitations.
Another interesting problem is the refinement of unresolved structures of experimental spectra that are due to the complex geometry and chemical composition of molecular magnets. Still another complication for the theoretical description is two-fold nature of the magnetic excitations in such systems. While on the molecular level these systems are an assembly of weakly-interacting spins and one can use a single-large-spin anisotropic Hamiltonian to reproduce the experimental data, each molecule is a complex system of strongly interacting atomic spins, which requires an accurate definition of the magnetic interactions and solution of the corresponding magnetic models.

All these problems are revealed in the case of Mn$_{12}$ (Ref.\onlinecite{Lis})  that is a popular system for molecular spintronics.
The theoretical investigations \cite{ourold, Bouk2007, Zeng} based on the density functional theory (DFT) numerical methods gave a correct description of the electronic and magnetic ground state properties such as energy gap, magnetic moment values of both the individual atomic and molecule spins. To describe the magnetic excitations in Mn$_{12}$ the single-molecule-spin Hamiltonian with uniaxial anisotropy was initially used. \cite{Friedman} Then additional couplings such as fourth-order transverse molecular anisotropy, spin-phonon interactions, etc. were artificially introduced into the model \cite{Sessoli} to simulate the tunneling effects. Despite of the fact that such a model approach reproduces the main features of the Mn$_{12}$ magnetic spectra, the underlying microscopic mechanisms are still unknown. The latter can be addressed by using numerical calculations based on DFT.

On the intra-molecular level there are single-ion anisotropy, inter-atomic isotropic and anisotropic exchange interactions which define  the main features of the excitation spectra. The previous theoretical investigations devoted to the Mn$_{12}$ system were mainly focused on the definition of the isotropic magnetic couplings between manganese atoms. \cite{ourold} Much less studied are anisotropic couplings such as Dzyaloshinskii-Moriya (DM) interaction and single-ion anisotropy. It was shown that they are important for explanation of magnetic excitations observed in inelastic neutron scattering measurements. \cite{8-spin} In turn, the authors of Ref.\onlinecite{Raedt}, proposed a decisive role of DM interactions in tunneling processes. However, the main problem is that a rather accurate knowledge of the Hamiltonian parameters is needed for the excitation spectra simulations.

The Hilbert space dimension of the realistic Mn$_{12}$ Hamiltonian is the most serious limitation for the theoretical consideration of the magnetic excitations and reproducing the experimental spectra. Normally, one exploits different symmetries of the system, so that the Hilbert space can be partitioned into sectors and the Hamiltonian matrix becomes block diagonal. \cite{Manmana} There could be lattice symmetries and/or spin symmetries, for instance the conservation of the $z$ projection of the spin. In the case of Mn$_{12}$ molecule we deal with the zero-dimensional object having a complex network of the Dzyloshinskii-Moriya interactions that mix the sectors with different total spins. Thus one faces the eigenvalue problem for the matrix of 10$^8 \times$ 10$^8$ to perform a realistic simulation of the magnetic excitations of the molecular magnet.  In such a situation the using of the simplified 8-spin Hamiltonian \cite{8-spin, Raedt} for description of Mn$_{12}$ was a demonstration of the computational hardware and software limits in the beginning of this century. Thanks to the constant development of distributed and shared memory computing systems, the realistic simulations  of the molecular magnets are available by means of realization of high-performance parallel algorithms.

The aim of our investigation is to develop a microscopic theory of the molecular magnetism in Mn$_{12}$, that describes the magnetic properties of the Mn$_{12}$ molecule from first-principles without any fitting procedure. It combines the local density approximation in conjunction the Hubbard $U$ term (LDA+$U$) for description of the ground state properties, the Green's function method for calculating the magnetic model parameters, and high-performance parallel exact diagonalization (ED) solver for simulating the experimentally observed spectra of the system. We consider such an approach to be preferable in comparison with previous ones, strongly relying on the fitting of the existing experimental data.  The latter often leads to the situation in which the same experimental curve can be fitted with completely different sets of parameters. Moreover, the microscopic mechanisms leading to realization of a particular magnetic structure remain unknown.

Our calculations for molecular magnet Mn$_{12}$ reveal the magnetic ordering that is richer than it was thought for almost 20 years. There are non-collinear patterns in the magnetic structure that are due to antisymmetric anisotropic exchange interaction between manganese atoms. For instance, there is a weak antiferromagnetic ordering for $z$-oriented magnetic structure. On the other hand, if the magnetic moments of manganese atoms are in the $xy$ plane, they cant from ferrimagnetic state in a similar way to antiferromagnets with weak ferromagnetism. To define the role of the inter-atomic interactions on the molecular level we use a Weiss-molecular-field-type approach.

 A consistent interpretation of the magnetic excitations in Mn$_{12}$ presented in the previous theoretical works is mainly based on the fitting of the inelastic neutron scattering (INS) spectra with simplified spin models. \cite{INS} Here we revise the  INS excitations by using the exact diagonalization of the full Mn$_{12}$ molecule Hamiltonian with parameters determined from the first-principles calculations.

\section{Computational methods}
{\it Electronic structure.} The Projector Augmented Wave (PAW) method as implemented in the VASP program package~\cite{vasp-paw} was employed to obtain an accurate description of the electronic and magnetic structure of Mn$_{12}$.
The spin-orbit coupling was taken into account within a non-collinear realization of the PAW method.~\cite{vasp-noncoll}
Correlation effects between Mn $d$ states were treated on a mean-field level using rotationally invariant LSDA+$U$ by Dudarev \textit{et al.}.~\cite{ldau-dudarev}
The plane wave energy cut-off of 600 eV was used along with a 4 $\times$ 4 $\times$ 4 k-point grid.

We also used the tight-binding linear-muffin-tin-orbital atomic sphere approximation (TB-LMTO-ASA) method. \cite{OKA} The exchange and correlation effects have been taken into account by using the LDA+$U$\cite{Anisimov} approach.

{\it Magnetic interactions.} In order to describe the magnetic excitations of Mn$_{12}$ we use the following spin Hamiltonian:
\begin{eqnarray}
\hat H =\sum_{ij} J_{ij} \hat {\vec S}_{i} \hat{\vec S}_{j} + \sum_{i \mu \nu} \hat {S}_{i}^{\mu} A^{\mu \nu}_{i} \hat {S}_{i}^{\nu} + \sum_{ij} \vec D_{ij} [\hat{\vec S}_{i} \times \hat {\vec S}_{j}] ,
\end{eqnarray}
where $J_{ij}$ is the isotropic exchange interaction, $\vec D_{ij}$ is Dzyaloshinskii-Moriya interaction and $A^{\mu \nu}_{i}$ is the element of the single-ion magnetic anisotropy tensor ($\mu, \nu = x,y,z$). The summation for inter-atomic couplings runs twice over every pair. Such a Hamiltonian contains the different combinations of the spin operators that conserve or do not conserve the total spin of the system $\mathcal {S}$. The combination of the first type are: $\hat {S}^z_i \hat {S}^z_i$, $\hat {S}^z_i \hat {S}^z_j$  and $\hat {S}^x_i \hat {S}^y_j - \hat {S}^y_i \hat {S}^x_j$. In turn, the following operators couple the levels with different total spin: $\hat {S}^x_i \hat {S}^z_j - \hat {S}^z_i \hat {S}^x_j$ ($\delta \mathcal{S} = \pm 1$) corresponding to the Dzyaloshinskii-Moriya interaction or $\hat {S}^x_i \hat {S}^y_i + \hat {S}^y_i \hat {S}^x_i$ ($\delta \mathcal{S} = \pm 2$) describing the single-ion anisotropy.

The main goal of our investigation is to define the parameters  of the spin Hamiltonian Eq.(1).
 According to the magnetic force theorem \cite{Lichtenstein} the variation of the total energy of the system due to a magnetic excitation can be expressed through the variation of the single-particle energy
\begin{eqnarray}
\delta E = - \int_{-\infty}^{E_{F}} \, d \epsilon \, \delta N (\epsilon),
\end{eqnarray}
here $N(\epsilon)$ is the integrated density of the electron state and $E_{F}$ is the Fermi energy.
Usually, the magnetic excitations related to a small rotation of the magnetic moments of the transition metal atoms from the collinear ground state are considered.
In this case the first and the second variations of the total energy written in the basis $|ilm\sigma\rangle$ (where $i$ denotes the site, $l$ the orbital quantum number, $m$- magnetic quantum number and
$\sigma$- spin index) are given by the following expressions
\begin{eqnarray}
\delta E=  - \frac{1}{\pi} \, \sum_{i} \int_{-\infty}^{E_{F}} d \epsilon \, {\rm Im} \, {\rm Tr}_{m,\sigma} \,  (\delta H_{i} \, G_{ii}) \label{firstvarE}
\end{eqnarray}
and
\begin{eqnarray}
\delta^{2} E=  - \frac{1}{\pi} \, \int_{-\infty}^{E_{F}} d \epsilon \, {\rm Im} \, {\rm Tr}_{m, \sigma} \, ( \sum_{i} \delta^{2} H_{i} \, G_{ii} \, \nonumber \\
+ \, \sum_{ij} \delta H_{i} \, G_{ij} \, \delta H_{j} \, G_{ji}).
\end{eqnarray}
Here $\delta H$ is the variation of the Hamiltonian, $G_{ii}$ and $G_{ij}$ are one-site and inter-site atomic Green's functions that can be calculated by using LDA+U approach.

Depending on the kind of the magnetic excitations we can define different parameters of the spin Hamiltonian for the atomic system. For instance, if the variation $\delta H_{i}$ is related to the rotation of the magnetic moments from the collinear ground state, then one can obtain the isotropic exchange interaction \cite{Lichtenstein}
\begin{eqnarray}
\label{J_Gr}
J_{ij} = -\frac{1}{4\pi S_i S_j} \int_{-\infty}^{E_{F}} d\epsilon \, {\rm Im}
{\rm Tr}_{m} (\Delta_{i} \,
G_{ij \, \downarrow} \, \Delta_{j} \, G_{ji \, \uparrow}),
\end{eqnarray}
where $\Delta_{i}$ is the magnetic splitting of the on-site potential and $S$ is the atomic spin.

In $3d$ systems, the spin-orbit coupling (SOC) in itself can be also considered as a perturbation. \cite{Solovyev, Bruno} In this case one can compute the magnetic anisotropy energy as
 \begin{eqnarray}
E_{anis} = -\frac{1}{2\pi} \sum_{ij} \int_{-\infty}^{E_{F}} d\epsilon \, {\rm Im}
 {\rm Tr}_{m, \sigma} (H^{so}_{i} \, G_{ij} \, H^{so}_{j} \, G_{ji}),
\end{eqnarray}
where $H^{so}_i = \lambda \vec L_i \vec S_i$ is the SOC operator for site $i$ ($\lambda$ = 0.05 eV). Changing the direction of the spin magnetization one can define all the elements of the $A^{\mu \nu}_{i}$ tensor.

There can be a mixed perturbation scheme with respect to the rotation and spin-orbit coupling, which leads to the antisymmetric anisotropic (Dzyaloshinskii-Moriya) exchange interaction \cite{MnCuN}
\begin{eqnarray}
\label{Dz}
D^{z}_{ij} = - \frac{1}{8 \pi S_i S_j} \, {\rm Re}
\int_{-\infty}^{E_{F}} d \epsilon \, \sum_{k}
\nonumber \\
\times {\rm Tr}_{m} (\Delta_{i} G_{ik}^{\downarrow} H^{so}_{k \, \downarrow \downarrow} G_{kj}^{\downarrow} \Delta
_{j} G_{ji}^{\uparrow} -
\Delta_{i} G_{ik}^{\uparrow} H^{so}_{k \, \uparrow \uparrow} G_{kj}^{\uparrow} \Delta_{j} G_{ji}^{\downarrow}
\nonumber \\
+ \Delta_{i} G_{ij}^{\downarrow} \Delta_{j} G_{jk}^{\uparrow} H^{so}_{k \, \uparrow \uparrow} G_{ki}^{\uparrow} -
\Delta_{i} G_{ij}^{\uparrow} \Delta_{j} G_{jk}^{\downarrow} H^{so}_{k \, \downarrow \downarrow} G_{ki}^{\downarrow}).
\end{eqnarray}

{\it Lanczos procedure.}  Once the parameters of the spin Hamiltonian $\hat{H}$ have been computed,
we apply the parallel Lanczos algorithm for shared memory systems to calculate the spin excitation spectrum. The main peculiarities of the Lanczos method we use are the following. Generating the Lanczos vector we do not store the Hamiltonian matrix elements, it is so-called diagonalization on the fly.
All the Lanczos vectors are stored on hard disk to perform their orthogonalization by the modified Gram-Schmidt method. \cite{Golub}
Our computational scheme is sensitive to the required number of eigenvalues and the number of Lanczos iterations.
In case of the Mn$_{12}$ molecule the calculation of 50 eigenvalues with good accuracy,  as measured by the variances of the energy in each of the approximate eigenstates,  requires about 1 Tb of disk space.
As we will show below such a number of eigenvalues of the full Mn$_{12}$ Hamiltonian can be defined as a minimum threshold for performing a realistic description of the experimental INS spectra.

\section{LDA+$U$ results}
{\it Electronic properties.} The first step of our investigation is to perform the ab-initio calculations for correct description of the ground state properties of the Mn$_{12}$ system. \cite{structure} For these purposes we have used the LDA+$U$ method. \cite{Anisimov} For LDA+$U$ calculations one needs to specify the values of the on-site Coulomb and intra-atomic exchange interactions.  The choice of $U$ and $J_H$ parameters for Mn$_{12}$ was extensively discussed by some of us in a prior work \cite{ourold}, where the values of $U$ parameter in the range between 4 and 8 eV were probed. These calculations have revealed a weak dependence of the magnetic quantities, such as magnetic moments and exchange integrals, on the $U$-parameter.  

Here we use $U$= 4 eV and $J_{H}$ = 0.9 eV for which the calculated value of the energy gap is close to the experimental one \cite{optics}. As we will show below, the obtained exchange integrals and magnetic anisotropy lead to the excitation spectrum that is in agreement with neutron scattering experiments. 
The values of the magnetic moments that are 2.8 $\mu_{B}$ for Mn$^{4+}$ and 3.7 $\mu_{B}$ for Mn$^{3+}$ agree well with previous theoretical results. \cite{ourold, Bouk2007, Zeng}

{\it Isotropic exchange interactions}. Previous works devoted to magnetic properties of Mn$_{12}$ are mainly based on the analysis of the spin Hamiltonian containing only the isotropic exchange interactions between manganese atoms. \cite{ourold} They define the largest energy scale for magnetic interactions and yield the ground state with $\mathcal{S}$=10 for the whole molecule. By using the eigenvectors and eigenvalues of the electronic Hamiltonian in the LDA+U approximation we calculated the full set of the isotropic exchange interactions employing Eq.(\ref{J_Gr}).

The comparison of the computed interactions with results of the previous theoretical and experimental works is presented in Table I. One can see that we have obtained a more detailed picture of magnetic couplings than before. For instance, the interactions corresponding to the bonds 1-9 (3.44 \AA) and 1-11 (3.45 \AA) are inequivalent in contrast to the prior studies.

\begin{table}[!h]
\centering
\caption [Bset]{Intra-molecular isotropic exchange interaction parameters (in meV) calculated by using LDA+$U$ approach. Positive sign corresponds to the antiferromagnetic coupling.}
\label {JDMI}
\begin {tabular}{cccccccc}
  \hline
  Bond (i,j) &  1-6 &  1-11 & 1-9 & 6-9 & 7-9 & 1-4 & 1-3 \\
  \hline
J$_{ij}$ (this work) & 4.6 & 1.0  & 1.7 & -0.45 & -0.37 & -1.55 & -0.5 \\
J$_{ij}$ (Ref.\onlinecite{ourold}) & 4.8 & 1.37  & 1.37 & -0.5 & -0.5 & -1.6 & -0.7 \\
J$_{ij}$ (Ref.\onlinecite{barbara}) & 7.4 & 1.72  & 1.72 & - & - & -1.98 & - \\
\hline
\end {tabular}
\end {table}

The method we use allows us to determine the orbital-resolved contributions to total exchange interaction between magnetic moments, $J_{ij} =\sum_{mm'} J^{mm'}_{ij}$, $m$ numerates $3d$ states of Mn atom.  Since we consider superexchange excitations through oxygen states, the individual $J_{ij}^{mm'}$ can be originated from two main microscopic mechanisms. The first one is antiferromagnetic kinetic Anderson exchange interaction\cite{Anderson} $J_{ij}^{mm'}=\frac{2 (t_{ij}^{mm'})^2}{U}$ that is due to the hopping processes between half-filled $3d$ orbitals. The second ferromagnetic mechanism results from the overlap of the half-filled and empty $3d$ orbitals of manganese atoms, $J_{ij}^{mm'} = - \frac{2 (t_{ij}^{mm'})^2 J_{H}}{U(U-J_{H})}$. \cite{Na2V3O7}

The orbital analysis of the calculated exchange integrals shows that the ferromagnetic interactions between Mn$^{4+}$ ions are purely of the second type. At the same time the interactions between  Mn$^{3+}$ and Mn$^{4+}$ ions are the result of a competition of antiferromagnetic and ferromagnetic contributions. For instance, this is the case for couplings 1-11 and 1-9 where a small difference in Mn-O-Mn bond angle and distance leads to a considerable difference in ferromagnetic contributions to the total exchange interaction.

\begin{figure}[!h]
\centering
\includegraphics[angle=90,width=90mm]{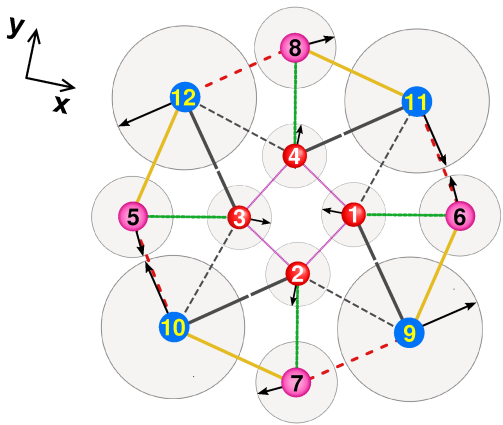}
\caption {Schematic representation of the atomic structure of Mn$_{12}$ molecule ($xy$ projection). Symmetry of magnetic interactions in Mn$_{12}$. Mn atoms are shown as spheres. Bonds of the same colour (style) can be transformed to each other by applying $\mathcal{S}_{4}$ symmetry operation. For instance, $\vec D_{6-9} \rightarrow  \vec D_{7-10} \rightarrow \vec D_{5-12} \rightarrow \vec D_{8-11}$. The arrows, locked inside transparent circles, show the transverse components of the magnetic moments away from the $z$ axis as comes out from the first-principles calculations. The actual ratio between the largest and smallest radii of the circles is supposed to be about 6 (see Table V for numerical data), which is reduced on the figure for better visualisation.
}
\end{figure}

{\it Anisotropic exchange interactions.} Having analyzed the isotropic exchange interaction between magnetic moments of manganese atoms we are going to discuss the antisymmetric anisotropic exchange interaction that can lead to non-collinear ground state of the Mn$_{12}$ system.
The calculated DM interaction parameters are presented in Table \ref{JDMI}. One can see that the absolute values of some individual DM interactions (1-11, 3-10, 2-9 and 4-12) are two order of magnitude smaller than the corresponding isotropic exchange integrals. It is about 10 times larger than one usually observes in transition metal crystals. \cite{Mazurenko}

According to the Neumann's principle \cite{Neumann} the DM interactions posses the symmetry of the crystal. In case of the Mn$_{12}$ system they are related by the symmetry of $\mathcal{S}_{4}$ (four-fold rotary-reflection axis parallel to $z$-direction) group. Taking into account that the Dzyaloshinskii-Moriya vector is an axial vector, one can obtain the following relation: $\vec D_{12} = (D^{x}, D^{y}, D^{z})$ transforms into $\vec D_{41} = (D^{y}, -D^{x}, D^{z})$. The same relation is valid for other bonds denoted  by the same color (Fig.~1).

Despite of the complex distorted geometry of the magnetic core of the Mn$_{12}$ molecule the DM vector symmetry for some bonds can be confirmed by the Moriya's rules. \cite{Moriya} There is the $\mathcal{C}_{2}$ rotational axis that is along $z$ direction and passes through the point bisecting the straight line between Mn1 and Mn3 atoms. It means that $z$ component of $\vec D_{13}$ is equal to zero.

The anisotropic exchange interactions in transition metal oxide can affect the magnetic structure in different ways. For instance, it can result in weak ferro- or antiferromagnetism, spin spiral state or others. \cite{Cheong}  As we will show below depending on the direction of the molecule magnetic moment, that can be controlled by an external magnetic field, the atomic spins deviate from the ferrimagnetic state in different planes. For instance, if $\vec B_{ext} || z$ the magnetic moments of Mn1, Mn3, Mn5  and Mn6 atoms will mainly cant in $xz$-plane. These perturbation theory results are confirmed by the LDA+$U$+SO calculations described below.

{\it  Molecular torque.}
It is important to define the effect of the inter-atomic DMI on the molecular level. Since the ground state of Mn$_{12}$ with canting of the molecule spin will be described below by means of first-principles LDA+$U$+SO calculations, here we would like to give a simplest and preliminary microscopic description of such an effect. 

Due to the symmetry restrictions only $z$-components of the Dzyaloshinskii-Moriya interactions contribute to the total magnetic torque of the whole molecule. It means that the canting of the molecule spin exists when the atomic spins are in the $xy$ plane.  Let us consider the ferrimagnetic ordering along $x$ axis, $\vec S_i = (\frac{3}{2}, 0, 0)$ for $i$ = 1..4 and $\vec S_{i} = (-2, 0, 0)$ for $i$ = 5..12. The canting of the molecule spin is formed by the canting of the individual atomic spins. To describe the latter in the simplest way we consider the independent excitations when the only one spin deviates from the ferrimagnetic ordering. It means that for each excitation the total energy of the system has the following dependence on the deviation angle 
\begin{eqnarray}
\Delta E =  - S^x_{i} \delta \phi^z_{i} \sum_{j} D^z_{ij} S^x_{j} - \frac{1}{2} S^x_{i} (\delta \phi^z_{i})^2 \sum_{j} J_{ij} S^x_{j},
\end{eqnarray}
where $\delta \phi^{z}_{i}$ is the rotational angle of the $i$th spin around $z$ axis.  
Thus the angle corresponding to the minimum of the energy is written in the form 

\begin{eqnarray}
\delta \phi^z_{i} = - \frac{\sum_{j} D^z_{ij}S^x_{j}}{\sum_{j} J_{ij} S^x_{j}}.
\end{eqnarray}      

We obtain $\delta \phi_{i}^z$ = 2.7 $\times$ 10$^{-3}$ for $i$=1..4,  $\delta \phi_{i}^z$ = -3.5 $\times$ 10$^{-4}$  for $i$=5..8 and $\delta \phi_{i}^z$ = -6.6 $\times$ 10$^{-3}$ for $i$=9..12. 
One can see that the sign of the deviation depends on the magnetic sublattice we consider. The magnetic moments of Mn$^{3+}$ and Mn$^{4+}$ ions have different orientations. It is similar to antiferromagnets with weak ferromagnetism. \cite{Mazurenko}   

The total molecule canting can be estimated as a sum of the individual atomic deviations, 

\begin{eqnarray}
\delta \phi_{mol}^z \approx \frac{1}{\mathcal{S}} \sum_{i} \delta \phi_{i}^z S^x_{i},
\end{eqnarray}
where the molecular spin $\mathcal{S}$=10. The obtained angle of 0.007 is in reasonable agreement with the LDA+$U$+SO result of  0.002.

We would like to stress that the canting of the individual atomic and molecular spins can be also realized through the non-diagonal elements of the single-ion anisotropy tensor. \cite{YMnO3} Such a scenario is considered below.

\begin{table}[!h]
\centering
\caption [Bset]{Intra-molecular anisotropic exchange interaction parameters calculated by using LDA+$U$ approach. $\vec R_{ij}$ is a radius vector connecting i$th$ and j$th$ atoms (in units of a=17.31 \AA).}
\label {JDMI}
\begin {tabular}{ccc}
  \hline
Bond (i,j) &$\vec R_{ij}$    &  $\vec D_{ij}$ (meV)   \\
  \hline
2-7  & (0.03; -0.16; 0.0)  &  ( -0.008 ; -0.013 ; -0.002 ) \\
4-8 & (-0.03; 0.16;  0.0)   &  (0.008 ; 0.013 ; -0.002 ) \\
1-6 & (0.16; 0.03; 0.0)  &  ( -0.013 ; 0.008 ; -0.002 ) \\
3-5 & (-0.16;  -0.03; 0.0)   &  (0.013 ; -0.008 ; -0.002 ) \\
  \hline
1-11 & (0.06; 0.18; 0.07)   &  ( -0.020; 0.03 ; -0.055 ) \\
3-10 & (-0.06;  -0.18;  0.07)   &  (0.020 ; -0.03 ; -0.055 ) \\
2-9 & (0.18;  -0.06;  -0.07)   &  (-0.03 ; -0.020 ; -0.055 ) \\
4-12 & (-0.18;  0.06;  -0.07)   &   (0.03 ; 0.020 ; -0.055 ) \\
  \hline
1-9  & (0.11; -0.16; 0.04)   &  ( 0.020; 0.014 ; 0.03) \\
3-12 & (-0.11;  0.16; 0.04)   &  ( -0.020 ; -0.014; 0.03) \\
2-10  & (-0.16; -0.11; -0.04)  &  ( -0.014 ; 0.020 ; 0.03) \\
4-11 & (0.16; 0.11; -0.04)   &  ( 0.014 ; -0.020 ; 0.03) \\
  \hline
6-9  & (-0.04; -0.18; 0.04)   &  ( -0.006 ; -0.004 ; -0.012) \\
5-12 & (0.04; 0.18; 0.04)  &  ( 0.006 ; 0.004 ; -0.012) \\
7-10 & (-0.18; 0.04; -0.04)   &  (0.004; -0.006; -0.012) \\
8-11 & (0.18;  -0.04;  -0.04)   &  (-0.004; 0.006; -0.012) \\
  \hline
7-9 & (0.15; 0.1; -0.07)   &   ( 0.020; -0.004 ; 0.012 ) \\
8-12 &(-0.15; -0.1; -0.07)   &   ( -0.020; 0.004; 0.012 ) \\
6-11 & (-0.1; 0.15; 0.07)   &  (-0.004 ; -0.020; 0.012) \\
5-10 & (0.1; -0.15; 0.07)   &  (0.004 ; 0.020; 0.012) \\
  \hline
4-1  & (-0.10; 0.06; 0.11 )  &    ( -0.014; 0.005; -0.013) \\
1-2  & (-0.06; -0.10; 0.11)  &    ( -0.005; -0.014; -0.013) \\
3-4  & (0.07; 0.1; 0.11)  &    ( 0.005; 0.014; -0.013) \\
2-3  & (-0.10; 0.07;  -0.11)  &    ( 0.014; -0.005; -0.013) \\
  \hline
1-3  & (-0.16; -0.03; 0.0)  &  ( -0.006; 0.030; 0 ) \\
2-4  & (-0.04;   0.17; 0.0)  &  ( -0.030; -0.006; 0 ) \\
  \hline
\end {tabular}
\end {table}

{\it Single-ion anisotropy.}
To calculate the magnetocrystalline anisotropy tensors for manganese atoms we have used the method proposed by Solovyev $et$ $al.$ \cite{Solovyev} These results are presented in Table III.
One can see that the smallest in-plane anisotropy of $A^{zz}_i - A^{xx}_i$=0.02 meV is observed for Mn1-Mn4 atoms. It is due to the fact that all the Mn-O bonds with the MnO octahedra are close in distances varying from 1.85 \AA \, to 1.91\AA. In turn the distortion of the Mn$^{3+}$ octahedra is much stronger  1.88 -2.25 \AA (Mn5-Mn8) and  1.89 - 2.18 \AA (Mn9-Mn12). That leads to a considerable difference between in-plane and out-of-plane anisotropies.
\begin{table}[!h]
\centering
\caption [Bset]{The elements of single-ion magnetic anisotropy tensors (in meV) obtained by using Green's function method, Eq.(6).}
\label {basisset}
\begin {tabular}{cc}
  \hline
Mn atom   & Anisotropy Tensor    \\
  \hline
    & 0.006 0.004  -0.002   \\
Mn1 & 0.004 -0.012 -0.001   \\
    & -0.002 -0.001  0.006    \\
  \hline
    & -0.012 -0.004  0.001   \\
Mn2 & -0.004 0.006 -0.002   \\
    & 0.001 -0.002  0.006    \\
  \hline
    & 0.006 0.004  0.002   \\
Mn3 & 0.004 -0.012 0.001  \\
    & 0.002 0.001 0.006    \\
  \hline
    & -0.012 -0.004 -0.001   \\
Mn4 & -0.004 0.006 0.002  \\
    & -0.001 0.002 0.006    \\
  \hline
    & 0.033 0  0.018   \\
Mn5 & 0 0.037 0.001  \\
    & 0.018 0.001 -0.07    \\
  \hline
    & 0.033 0  -0.018   \\
Mn6 & 0 0.037 -0.001   \\
    & -0.018 -0.001  -0.07    \\
  \hline
    & 0.037  0 0.001     \\
Mn7 & 0 0.033  -0.018  \\
    & 0.001   -0.018 -0.07    \\
  \hline
    & 0.037  0 -0.001     \\
Mn8 & 0 0.033  0.018  \\
    & -0.001   0.018 -0.07    \\
  \hline
    & 0.020  -0.015  -0.048   \\
Mn9 & -0.015 0.015 -0.028  \\
    & -0.048 -0.028 -0.035    \\
  \hline
    & 0.015  0.015 0.028   \\
Mn10 & 0.015 0.020 -0.048  \\
    & 0.028 -0.048 -0.035    \\
  \hline
    & 0.015  0.015 -0.028   \\
Mn11 & 0.015 0.020 0.048  \\
    & -0.028 0.048 -0.035    \\
  \hline
    & 0.020  -0.015 0.048   \\
Mn12 & -0.015 0.015 0.028  \\
    & 0.048 0.028 -0.035    \\
  \hline
\end {tabular}
\end {table}

Another important result is that there are strong non-diagonal elements of the single-ion anisotropy tensor for the Mn9-Mn12 atoms.
These elements provide an additional contribution to the spin moment canting from the $z$-oriented collinear configuration. For instance, the element A$^{xz}_{9}$ leads to the canting around $y$ axis in the $xz$ plane. According to the calculated single-ion anisotropy the largest deviations take place for Mn9-Mn12 atoms. These results will be confirmed in the framework of the LDA+$U$+SO calculations.

It is interesting to estimate the anisotropy of the whole molecule by summarizing the anisotropies of the individual atoms. We obtain the following molecular anisotropy tensor
\begin{eqnarray}
\mathcal{A}_{mol} = \frac{1}{\mathcal{S}^2} \sum _{i=1}^{12} A_{i} S_{i}^{2}= \left( \begin{array}{ccc}
0.008 & 0  & 0 \\
0 &  0.008 & 0  \\
0 &  0 &  -0.016
\end{array} \right),
\end{eqnarray}
where the molecular spin $\mathcal{S}$=10, the atomic spins $S_{i} = \frac{3}{2}$ (for Mn1-Mn4) and $S_{i}$ = 2 (for Mn5-Mn12). The easy axis is along $z$ direction and $xy$ plane is the hard plane.
The corresponding single-molecule anisotropy can be estimated $\mathcal{A}^{zz}_{mol} - \mathcal{A}^{xx}_{mol}$ = -0.28 K. That is in reasonable agreement with results of experimental fitting. It is important to note that the non-diagonal elements of $\mathcal{A}_{mol}$ are zero. Thus the molecule torque is fully provided by the Dzyaloshinskii-Moriya interaction. However, since we obtain the solution with $xy$ hard plane, all the magnetic configurations with in-plane molecular spin correspond to the same energy and there is no energy gain due to the canting of the molecular spin.

\section{LDA+$U$+SO results}
The results obtained by using the perturbation theory on rotation of the magnetic moments and spin-orbit coupling should be confirmed by a numerical approach taking into account the spin-orbit coupling in the electronic Hamiltonian.
\begin{table}[t]
\centering
\caption [Bset]{Ground state properties obtained from LDA+$U$+SO calculations. $E_{tot}$ (in meV) and $\vec M_{tot}$ (in $\mu_B$) are the calculated total energy and total magnetization of the Mn$_{12}$ molecule.}
\begin {tabular}{ccccc}
  \hline
Magnetization & $E_{tot}$  & $M_{tot}^{x}$  & $M_{tot}^{y}$  & $M_{tot}^{z}$  \\
  \hline
  $X$ &  6.67 &   -19.99 &    0.045 & 0.0    \\
  $Y$ &  6.67 &  -0.045  &  -19.99  &  0.0    \\
  $Z$ & 0 &  0.0  &   0.0 &  -19.99    \\
  \hline
\end {tabular}
\end {table}
For these purposes we performed the LDA+$U$+SO calculations \cite{vasp-noncoll} with different orientations of the total magnetization of the Mn$_{12}$ molecule (Table IV).  In all the cases the performed calculations revealed non-collinear ground state for atomic spins of the Mn$_{12}$ molecule. The $z$-oriented configuration corresponds to the minimum of the total energy.
The energy difference between $z$- and $x$-oriented states gives us opportunity to estimate the anisotropy of the molecule with $\mathcal{S}$=10, $\frac{E^{Z}_{tot} - E^{X}_{tot}}{\mathcal{S}^2}$ = - 0.90 K. This value is in reasonable agreement with experimental estimate of 0.56 K obtained in high-frequency EPR measurements. \cite{EPR}

We also observe the canting of the total magnetic moment of the molecule for $x$- and $y$-oriented configurations.  This effect is due to the inter-atomic Dzyaloshinskii-Moriya interactions. The canting angle can be estimated as $\delta \phi_{mol}$= 0.002, which is in a reasonable agreement with perturbation theory results.

\begin{table}[t]
\centering
\caption [Bset]{Individual site-resolved components of spin and orbital magnetic moments (in $\mu_{B}$) obtained from LDA+$U$+SO calculations for $\vec M_{tot} || z$.}
\begin {tabular}{ccccccc}
  \hline
Mn atom & $M_{S}^{x}$  & $M_{S}^{y}$  & $M_{S}^{z}$  & $M_{L}^{x}$  & $M_{L}^{y}$  & $M_{L}^{z}$  \\
  \hline
Mn1  & -0.002&  0.000 & 2.835  &  0.000 &  0.000 & -0.017 \\
Mn2  &  0.000& -0.002 & 2.835  &  0.000 &  0.000 & -0.017 \\
Mn3  &  0.002&  0.000 & 2.835  &  0.000 &  0.000 & -0.017 \\
Mn4  &  0.000&  0.002 & 2.835  &  0.000 &  0.000 & -0.017 \\
\hline
Mn5  &  0.005& -0.002 & -3.720 & -0.004 &  0.000 & 0.027 \\
Mn6  & -0.005&  0.002 & -3.720 &  0.004 &  0.000 & 0.027 \\
Mn7  & -0.002& -0.005 & -3.720 &  0.000 &  0.004 & 0.027 \\
Mn8  &  0.002&  0.005 & -3.720 &  0.000 & -0.004 & 0.027 \\
\hline
Mn9  & -0.002&  0.002 & -3.738 &  0.012 &  0.006 & 0.022 \\
Mn10 & -0.002& -0.002 & -3.738 & -0.006 &  0.012 & 0.022 \\
Mn11 &  0.002&  0.002 & -3.738 &  0.006 & -0.012 & 0.022 \\
Mn12 &  0.002& -0.002 & -3.738 & -0.012 & -0.006 & 0.022 \\
 \hline
\end {tabular}
\end {table}

Let us analyze the $z$-oriented configuration. The orientations of the spin and orbital magnetic moments are presented in Fig.1 and Table V. One can see that they obey the symmetry operations of the $\mathcal{S}_{4}$ group. We observe a weak in-plane antiferromagnetic ordering induced by the Dzyaloshinskii-Moriya interaction.  

In turn, for $x$-oriented magnetic structure (Table VI) there is no compensation of the $y$ components of the magnetic moments, which leads to the deviation of the molecule spin from the $x$ direction.  Thus the obtained LDA+$U$+SO results confirms our analysis of the anisotropic exchange interactions between manganese atoms (Section III).

\begin{table}[!h]
\centering
\caption [Bset]{Individual site-resolved components of spin and orbital magnetic moments (in $\mu_{B}$) obtained from LDA+$U$+SO calculations for $\vec M_{tot} || x$.}
\begin {tabular}{ccccccc}
  \hline
Mn atom & $M_{S}^{x}$  & $M_{S}^{y}$  & $M_{S}^{z}$  & $M_{L}^{x}$  & $M_{L}^{y}$  & $M_{L}^{z}$ \\
  \hline
Mn1 & 2.835 & 	0.007&	0.007&	-0.018&	0.000&	0.001\\
Mn2 & 2.834 &	0.009&	-0.002&	-0.019&	0.000&	0.000\\
Mn3 & 2.835 &      0.007&	-0.007&	-0.018&	0.000&	-0.001\\
Mn4 & 2.834 &	0.009&	0.002&	-0.019&	0.000&	0.000\\
\hline
Mn5 & -3.721 &	-0.005&	0.001&	0.005&	0.000&	-0.004\\
Mn6 & -3.721 &	-0.005&	0.000&	0.005&	0.000&	0.004\\
Mn7 & -3.721 &	-0.003&	0.005&	0.006&	0.000&	-0.001\\
Mn8 & -3.721 &	-0.004&	-0.004&	0.006&	0.000&	0.001\\
\hline
Mn9 & -3.738 &	0.002&	-0.026&	0.011&	0.004&	0.012\\
Mn10 & -3.738 &	0.015&	0.019&	0.013&	-0.004&	-0.007\\
Mn11 & -3.738 &	0.015&	-0.019&	0.013&	-0.004&	0.007\\
Mn12 & -3.738 &	0.002&	0.026&	0.011&	0.004&	-0.012\\
\hline 
\end {tabular}
\end {table}

\section{Exact diagonalization results}
Having analyzed the magnetic ground state of the Mn$_{12}$ system we are going to study quantum spin excitation spectrum.
For that, the constructed spin Hamiltonian Eq.(1) is solved by means of exact diagonalization approach. Our ED solver is based on the parallel implementation of the Lanczos algorithm and gives us opportunity to calculate 50 lowest eigenvalues and the corresponding eigenfunctions. It means one can simulate the magnetic properties of the Mn$_{12}$ at finite temperatures.

The diagonalization results are presented in Fig.2.  All the calculated energy levels correspond to the total spins  $\mathcal {S}= 10$ and $\mathcal {S}= 9$. There is a gap of about  50 K  between the states with different $\mathcal{S}$.  The splittings between the nearest levels of the $\mathcal{S}=10$ band are not uniform, they decrease with the energy increase.
\begin{figure}[!t]
\centering
\includegraphics[angle=0,width=90mm]{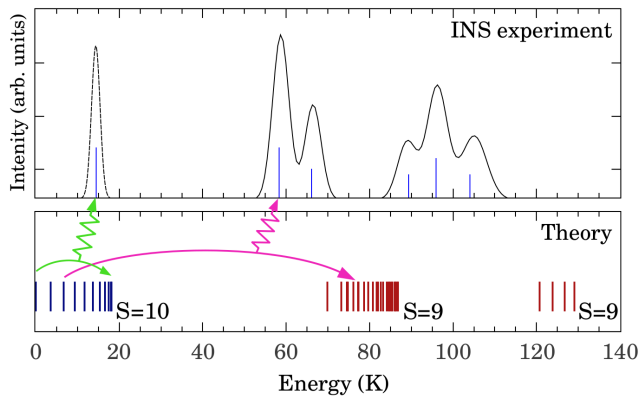}
\caption {Schematic comparison of the theoretical spectrum obtained by diagonalizing Eq.(1) and INS spectrum taken from Ref.\onlinecite{INS} (Fig.6 and 8 therein).  The arrows denote the intra- and inter-band transitions that correspond to the excitations observed in the INS experiment.}
\end{figure}

It is convenient to compare our low-energy spectrum with that measured in inelastic neutron scattering experiments (INS) \cite{INS} with selection rule $\delta \mathcal{S} = 0, \pm 1$. Based on the calculated eigenvalues we attribute the INS features at 14 K with the transition $\delta \mathcal{S} = 0$ and peaks at 57 (66 K) with transition to the levels with $\mathcal{S} = 9$. Such a excitation picture contradicts to the previous results obtained for the simplified Mn$_{12}$ models \cite{8-spin, INS} where the first excitation of 14.4 K was associated with the transition $\mathcal{S} = 10 \rightarrow  9$. Our simulations have shown that the account of the DMI couplings leads to an energy shift of the excited levels corresponding to $\mathcal{S} = 9$ and the structure of the $\mathcal{S} =  10$ band does not change.

\section{Conclusion}
In conclusion,  using the modern numerical techniques for calculating the magnetic interactions we propose a realistic spin model of the Mn$_{12}$ molecular magnet. Such a model contains the complete set of the isotropic and anisotropic magnetic interactions. Moreover, the parameters of the model take into account tiny details of the Mn$_{12}$ atomic structure. For instance, some bonds between manganese atoms have small differences in length and angle of the metal-oxygen-metal pathways and they were assumed to be equivalent in the previous theoretical investigations. In our work we show that such a small difference in geometry leads to a strong distinction in exchange interactions, which can be explained from a microscopic point of view by analyzing orbital contributions to the exchange integrals.

Our first-principles results provide very strong evidence of a complex non-collinear ordering of the manganese magnetic moments. It is caused by the Dzyaloshinskii-Moriya interactions and non-diagonal elements of the single-ion anisotropy whose symmetry fully obeys the $\mathcal {S}_4$ symmetry of the Mn$_{12}$ system. Similar non-collinear patterns due to DM interactions were recently found in famous itinerant magnet MnSi. \cite{Dmitrienko} The authors of the work proposed an approach to measure tilting components of the magnetic moment by using x-ray and neutron diffraction techniques that can be also used in case of the Mn$_{12}$ system.

An important part of our work is the estimation of the molecular anisotropic field, which is used to construct a microscopically justified molecular-single-spin Hamiltonian. The transition from individual atomic magnetic moments to macroscopic  magnetic moment of the Mn$_{12}$ molecule is confirmed by LDA+$U$+SO calculations.  In addition to the well-known easy axis that is along the $z$ direction we found the $xy$ hard plane.
Another interesting result is that the inter-atomic Dzyaloshinskii-Moriya interactions produce a torque acting on the molecular magnetic moment in the $xy$ plane, giving rise to a weak ferromagnetic component of the magnetization.

The exact diagonalization of the full spin Hamiltonian with parameters determined from first-principles calculations gives us an opportunity to provide a distinct classification of the INS peaks with respect to the spin transitions in the Mn$_{12}$ system. In contrast to previous considerations the low-energy excitations are due to the $\mathcal{S}=10$ intra-band transitions.

\section{Acknowledgements}
We thank I.V. Solovyev and D.W. Boukhvalov for helpful discussions and M.V. Valentyuk for technical assistance with LDA calculations. The work of VVM is supported by the Ministry of Education and Science of the Russian Federation, project 1751. MIK acknowledges a financial support from European Research Council, Advanced Grant 338957-FEMTO/NANO.


\begin{thebibliography}{99}

\bibitem{Bogani}
L. Bogani and W. Wernsdorfer,
Nature Materials {\bf 7}, 179 (2008).

\bibitem{Mannini}
M. Mannini, F. Pineider, P. Sainctavit, C. Danieli, E. Otero, C. Sciancalepore, A. M. Talarico, M. A. Arrio, A. Cornia, D. Gatteschi, and R. Sessoli, 
Nature Materials {\bf 8}, 194 (2009).

\bibitem{Lis}
T. Lis,  Acta Crystallogr., Sect. B: Struct Crystallogr. Cryst Chem. {\bf 36}, 2042 (1980).

\bibitem{ourold}
D.W. Boukhvalov, A. I. Lichtenstein, V. V. Dobrovitski, M. I. Katsnelson, B. N. Harmon, V. V. Mazurenko, and V. I. Anisimov, 
Phys. Rev B {\bf 65}, 184435 (2002).

\bibitem{Bouk2007}
D. W. Boukhvalov, M. Al-Saqer, E. Z. Kurmaev, A. Moewes, V. R. Galakhov, L. D. Finkelstein, S. Chiuzbaian, M. Neumann, V. V. Dobrovitski, M. I. Katsnelson, A. I. Lichtenstein, B. N. Harmon, K. Endo, J. M. North, and N. S. Dalal, 
Phys. Rev. B {\bf 75}, 014419 (2007).

\bibitem{Zeng}
Z. Zeng, Diana Guenzburger, and D. E. Ellis, 
Phys. Rev. B {\bf 59}, 6927 (1999).

\bibitem{Friedman}
J.R. Friedman, M.P. Sarachik, J. Tejada, and R. Ziolo, 
Phys. Rev. Lett. {\bf 76}, 3830 (1996).

\bibitem{Sessoli}
A. Fort, A. Rettori, J. Villain, D. Gatteschi and R. Sessoli, 
Phys. Rev. Lett. {\bf 80}, 612 (1998).

\bibitem{8-spin}
M.I. Katsnelson, V.V. Dobrovitski, and B.N. Harmon, 
Phys. Rev. B 59, 6919 (1999).

\bibitem{Raedt}
H. A. De Raedt, A. H. Hams, V. V. Dobrovitski, M. Al-Saqer, M. I. Katsnelson, and B. N. Harmon, J. Magn. Magn. Mater. 246, 392 (2002)

\bibitem{Manmana}
S.R. Manmana, {\em Nonequilibrium dynamics of strongly correlated quantum systems},
Ph.D. Thesis, Universit\"{a}t Stuttgart (2006).

\bibitem{INS}
M. Hennion, L. Pardi, I. Mirebeau, E. Suard, R. Sessoli, and A. Caneschi, 
Phys. Rev. B {\bf 56}, 8819 (1997).

\bibitem{vasp-paw} G. Kresse and D. Joubert, Phys. Rev. B {\bf 59}, 1758 (1999);
G. Kresse, M. Marsman and J. Furthm\"uller, {\em VASP: Vienna Ab-initio Simulation Package} (http://cms.mpi.univie.ac.at/VASP/).

\bibitem{vasp-noncoll} D. Hobbs, G. Kresse, and J. Hafner, Phys. Rev. B {\bf 62}, 11556 (2000).

\bibitem{ldau-dudarev} S. L. Dudarev, G. A. Botton, S. Y. Savrasov, C. J. Humphreys, and A. P. Sutton, Phys. Rev. B {\bf 57}, 1505 (1998).

\bibitem{OKA}
O.~K. Andersen, 
Phys. Rev. B {\bf 12}, 3060 (1975).

\bibitem{Anisimov}
V.~I. Anisimov, F. Aryasetiawan, and A.~I. Lichtenstein, J. Phys.: Condens. Matter,{\bf \ 9}, 767 (1997).

\bibitem{Lichtenstein} A. I. Liechtenstein, M. I. Katsnelson, V. P. Antropov, and
V. A. Gubanov, J. Magn. Magn. Mater. {\bf 67}, 65 (1987).

\bibitem{Solovyev}  I.V. Solovyev, P.H. Dederichs, and I. Mertig, Phys. Rev. B {\bf 52}, 13419 (1995).

\bibitem{Bruno}  P. Bruno, Phys. Rev. B {\bf 39}, 865 (1989).

\bibitem{MnCuN}
A.N. Rudenko, V.V. Mazurenko, V.I. Anisimov, and A.I. Lichtenstein, Phys. Rev. B {\bf 79}, 144418 (2009).

\bibitem{Golub}
G. H. Golub and C. F. Van Loan, {\em Matrix Computations},
John Hopkins University Press (1996).

\bibitem{structure}
The calculations were performed for the system Mn$_{12}$O$_{12}$(HCOO)$_{16}$(H$_{2}$O)$_{4}$ (Ref.\onlinecite{ourold}) that is a simplified version of the Mn$_{12}$ crystal structure reported by Lis \cite{Lis}. The changes mainly concern the methyl groups CH$_{3}$ that were replaced by hydrogen atoms.

\bibitem{optics}
J.M. North, D. Zipse, N. S. Dalal, E. S. Choi, E. Jobiliong, J. S. Brooks, and D. L. Eaton, 
Phys. Rev. B {\bf 67}, 174407 (2003).

\bibitem{Anderson}
P.W. Anderson,
Phys. Rev. {\bf 115,} 2 (1959);
Solid State Physics {\bf 14,} 99 (Academic, New York 1963).

\bibitem{barbara}
B. Barbara, D. Gatteschi, A. A. Mukhin, V. V. Platonov, A. I. Popov, A. M. Tatsenko, and A. K. Zvezdin, {\em Nano - scale ferrimagnet Mn12Ac in ultra-high magnetic field}, in Proceedings of Seventh International Conference on Megagauss Magnetic Field Generation and Related Topics, Sarov, 1996, 853 (1997).

\bibitem{Na2V3O7}
V. V. Mazurenko, F. Mila, V. I. Anisimov, 
Phys. Rev. B {\bf 73}, 014418 (2006).

\bibitem{Mazurenko}
V. V. Mazurenko and V. I. Anisimov, 
Phys. Rev. B {\bf 71}, 184434 (2005).

\bibitem{Neumann}
F. E. Neumann,  {\em Vorlesungen {\"u}ber die Theorie der Elastizit{\"a}t der festen K{\"o}rper und des Licht{\"a}thers}, edited by O. E. Meyer. Leipzig, B. G. Teubner-Verlag, 1885.

\bibitem{Moriya}
T. Moriya,  Phys. Rev. {\bf 120}, 91 (1960).

\bibitem{Cheong}
S. W. Cheong and M. Mostovoy, Nature Mater. 6, 13 (2007).


\bibitem{YMnO3}
I. V. Solovyev, M. V. Valentyuk, and V. V. Mazurenko, 
Phys. Rev. B {\bf 86}, 054407 (2012).

\bibitem{EPR}
A. L. Barra, D. Gatteschi, and R. Sessoli, Phys. Rev. B {\bf 56}, 8192 (1997).

\bibitem{Dmitrienko}
V. E. Dmitrienko and V. A. Chizhikov, Phys.Rev. Lett. {\bf 108}, 187203 (2012).

\end{thebibliography}
\end{document}